\documentclass[12pt]{article}
\pdfoutput=1
\usepackage[utf8]{inputenc}
\usepackage{amsmath}
\usepackage{amssymb}
\usepackage{units}
\usepackage{graphicx}
\usepackage{slashed}
\usepackage{array}
\usepackage{subcaption}
\usepackage[leftcaption,raggedright]{sidecap}
\usepackage{caption}
\usepackage[hidelinks]{hyperref}

\newcommand\pubnumber{NuPhys2023-Jasmine Simms}
\newcommand\pubdate{\today}

\def\napoli{University of Oxford}

\def\support{\footnote{
  Science and Technology Facilities Council.
}}

\def\Title#1{\begin{center} {\Large #1 } \end{center}}
\def\Author#1{\begin{center}{ \sc #1} \end{center}}
\def\Address#1{\begin{center}{ \it #1} \end{center}}

\newcommand\pubblock{\rightline{\begin{tabular}{l} \pubnumber\\
         \pubdate  \end{tabular}}}
\newenvironment{Abstract}{\begin{quotation}  }{\end{quotation}}
\newenvironment{Presented}{\begin{quotation} \begin{center} 
             PRESENTED AT\end{center}\bigskip 
      \begin{center}\begin{large}}{\end{large}\end{center} \end{quotation}}

\def\beq{\begin{equation}}
\def\eeq#1{\label{#1}\end{equation}}
\def\eeqn{\end{equation}}

\def\beqa{\begin{eqnarray}}
\def\eeqa#1{\label{#1}\end{eqnarray}}
\def\eeqan{\end{eqnarray}}

\let\bar=\overbar

\def\Dslash{\not{\hbox{\kern-4pt $D$}}}
\def\dslash{\not{\hbox{\kern-2pt $\del$}}}

\def\msb{{\bar{\ssstyle M \kern -1pt S}}}

\begin{document}
\begin{titlepage}
\pubblock

\vfill
\Title{Muon Track Reconstruction in the Scintillator Phase of SNO+}
\vfill
\Author{ Jasmine Simms\support}
\Address{\napoli}
\vfill
\begin{Abstract}
The large depth of the SNO+ experiment (2070 m, 6010 m.w.e.) means that only a few muons per day pass through the detector. However, their high energy causes muon induced backgrounds which can affect multiple physics analyses. Reconstructing the muon track would allow for improved rejection for these induced backgrounds. Currently there is no muon tracker for the scintillator phase of SNO+. This poster presents a novel method of muon track reconstruction by using the high photon sampling from muons and the assumption that each PMT first registers a hit from a photon that takes the fastest possible path from the muon entry point to the PMT.

\end{Abstract}
\vfill
\begin{Presented}
NuPhys2023, Prospects in Neutrino Physics\\
King's College, London, UK,\\ December 18--20, 2023
\end{Presented}
\vfill
\end{titlepage}
\def\thefootnote{\fnsymbol{footnote}}
\setcounter{footnote}{0}

\section{Introduction}
The SNO+ experiment is a liquid scintillator detector located at SNOLAB at a depth of 2.07 km (6.01 km water equivalent depth) which reduces the cosmic muon rate to 3 per hour [1]. High energy cosmic muons produce large quantities of scintillation light as well as causing spallation neutrons and isotopes, for example Carbon-11, which can be backgrounds for many physics analyses [2]. Muons are identified in the SNO+ detector using information including hits from outward looking PMTs, which face outwards into the cavity, as well as the total number of hit PMTs. A 20 s muon veto is applied after each identified muon to remove short half-life muon followers from data. This veto significantly reduces the livetime for already statistics limited analysis. 

Track reconstruction would allow for improved rejection of muon induced backgrounds and increased detector livetime by vetoing only a cylindrical volume around the track rather than the whole detector. It could also be used for improved analysis of muon followers. There is currently no muon track reconstruction for the scintillator phase of SNO+. A novel method of muon track reconstruction has been developed by making use of the high photon sampling from muons.

\section{Reconstruction Method}
Due to the high energy of the cosmic muons, there is sufficient quantity of scintillation light to hit all of the PMTs in SNO+ multiple times. Therefore, it is assumed that each PMT is hit by the photon that takes the fastest path between the entry point and the PMT itself. In Figure 1, the photon paths corresponding to the earliest hit time are shown for a hypothetical muon track. These paths are calculated using the relative speeds of the muon and the photons in the scintillator [3].
\begin{figure}[ht]
\centering
\includegraphics[width=.35\linewidth]{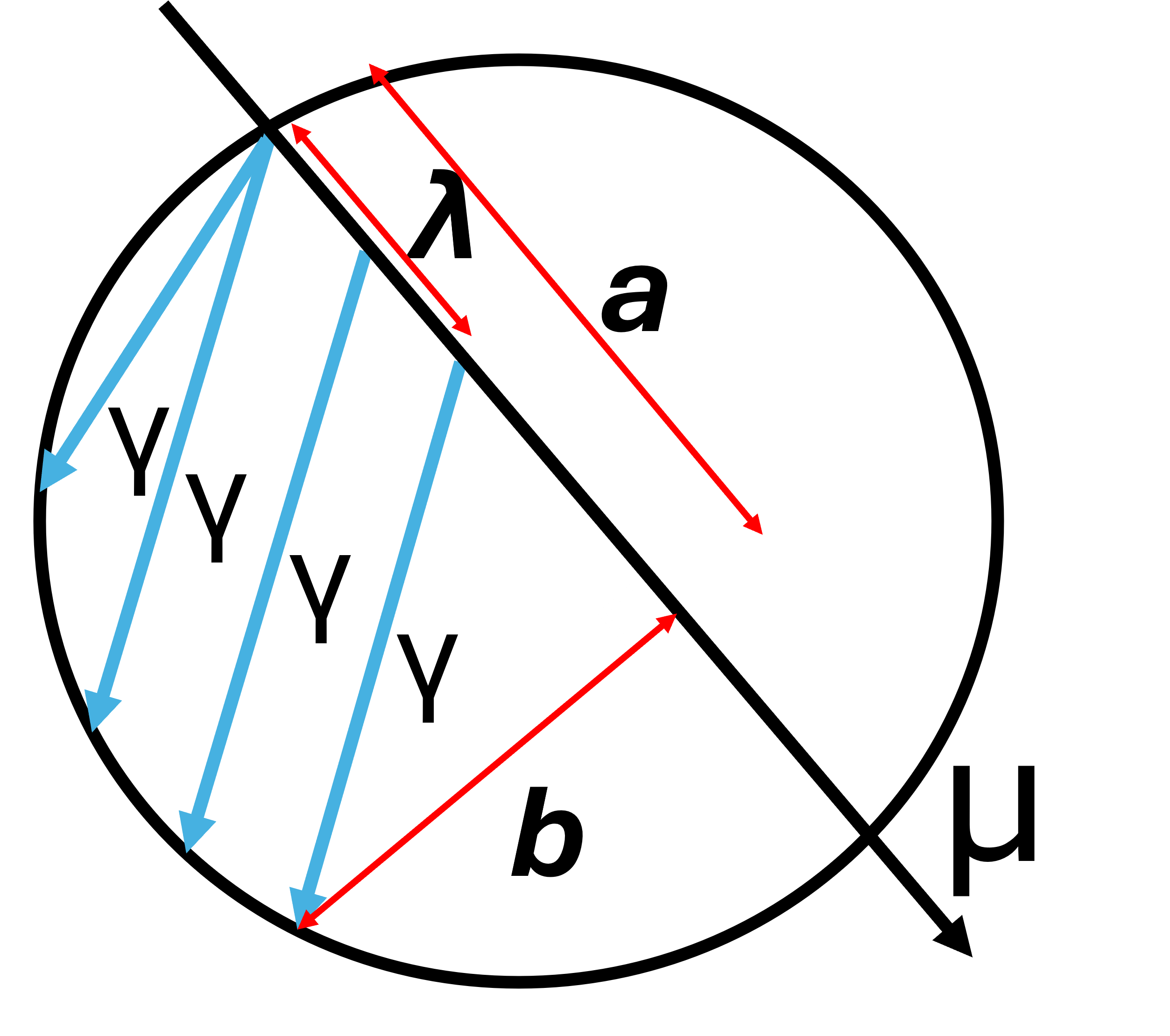}
\caption{Photon paths corresponding to the earliest hit time for an hypothetical muon track. $\lambda$ is the distance along the muon track between the photon creation point and the muon entry point. $a$ and $b$ are values used in Equation 1.}
\end{figure}

The time residual is defined as the difference between the time the photon was created and the time it triggers the detector.
\begin{equation}
    T_{res}=t_{hit}-\frac{n}{c}\sqrt{(a-\lambda)^2+b^2}
\end{equation}
Where $t_{hit}$ is the PMT hit time, $n$ is the refractive index of the scintillator and $c$ is the speed of light. $\lambda$, $a$ and $b$ are defined in Figure 1. The second term in Equation 1 is the transit time for each photon from the production point along the muon track to the hit PMT. 

$T_{res}$ and $\lambda$ are calculated using the position and hit time of each hit PMT for a sample muon track. Figure 2 shows the $T_{res} - \lambda$ plots which represents the distance travelled by the muon in the detector, along with the best fit line and its variance. The excess at $\lambda=0$ is due to the large number of photons assumed to have been generated at the muon entry point (see Figure 1). In order to avoid the fit being distorted by this excess, a cut at $\lambda>0.5$ is applied when calculating the slope. All point are included in the calculation of the variance.

\begin{figure}[ht]
\centering
\begin{subfigure}{.49\textwidth}
\centering
\includegraphics[width=.99\linewidth]{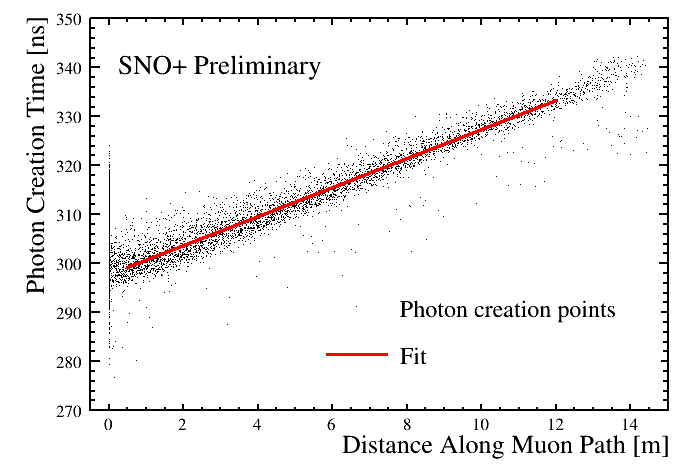}
\caption{True muon track}
\end{subfigure}
\begin{subfigure}{.49\textwidth}
\centering
\includegraphics[width=.99\linewidth]{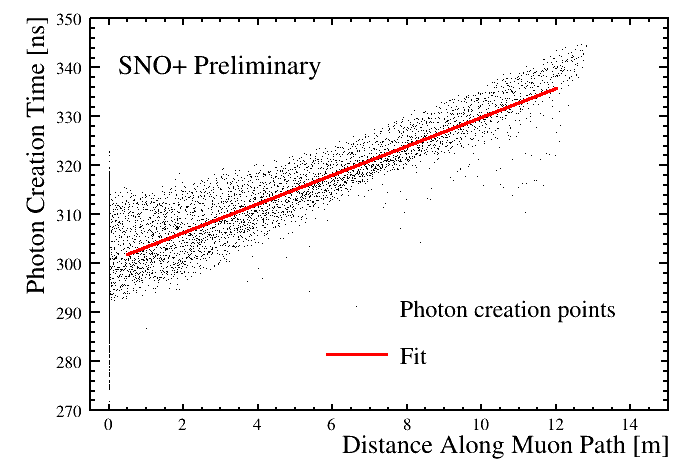}
\caption{Incorrect muon track}
\end{subfigure}
\begin{subfigure}{.49\textwidth}
\centering
\includegraphics[width=.99\linewidth]{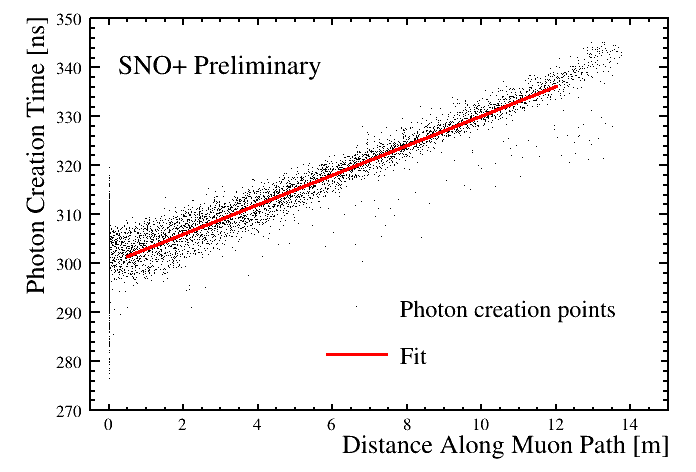}
\caption{Fitted muon track close to the true value}
\end{subfigure}
\caption{Graph of $T_{res}-\lambda$ as defined in Figure 1}
\end{figure}

Figure 2a shows the case when the true muon track is used and therefore shows a straight line with low variance, Figure 2b shows a high variance due to an incorrect trial muon track, whereas Figure 2c shows a much lower variance, indicating this trial muon track is close to the true track.

Muon tracks are defined by their entry position and direction. For a potential muon track the variance is calculated from the $T_{res} - \lambda$ graph and this variance is minimised using an adaptive grid search over the entry positions and directions followed by Minuit optimiser to find the reconstructed track.

\section{Angular Resolution}
The reconstruction method was tested on simulated cosmic muons in the SNO+ detector. A selection of these reconstructed tracks compared to the true tracks are shown in Figure 3. The fit is close to the true track for tracks that travel a significant distance through the detector. The fit is significantly worse for muons that graze the edge of the detector due to the short track length, however these are expected to be a minor contribution to the total SNO+ muons
\begin{figure}[ht]
\centering
\begin{subfigure}{.49\textwidth}
\centering
\includegraphics[width=.99\linewidth]{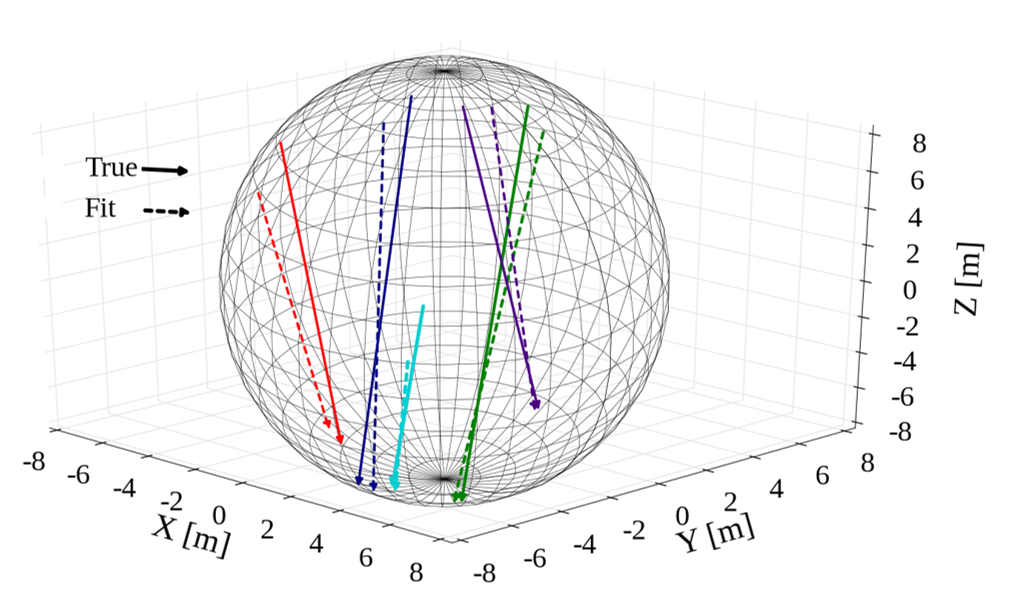}

\caption{Muons that travel significant distance in the detector}
\end{subfigure}
\begin{subfigure}{.49\textwidth}
\centering
\includegraphics[width=.99\linewidth]{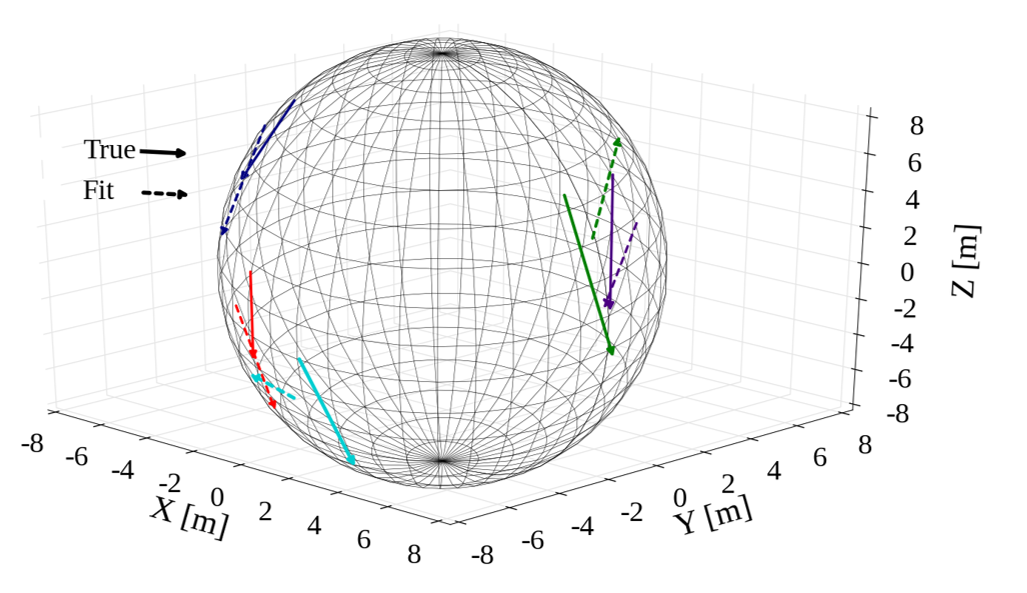}
\caption{Muons that graze the edge of the detector}
\end{subfigure}
\caption{True and reconstructed tracks for simulated cosmic muons }
\end{figure}
\begin{figure}[ht]
\centering
\begin{subfigure}{.49\textwidth}
\centering
\includegraphics[width=.99\linewidth]{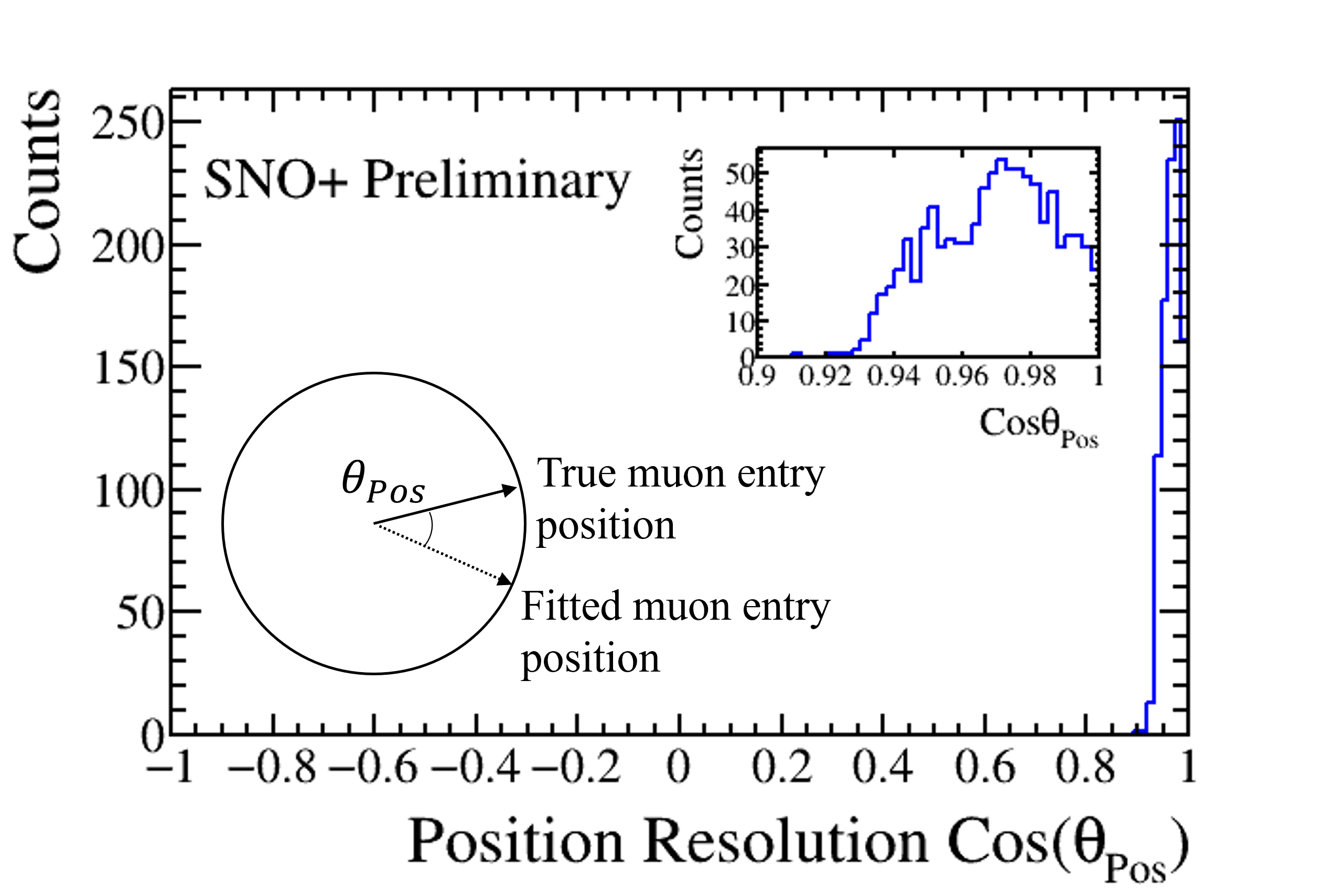}
\end{subfigure}
\begin{subfigure}{.49\textwidth}
\centering
\includegraphics[width=.99\linewidth]{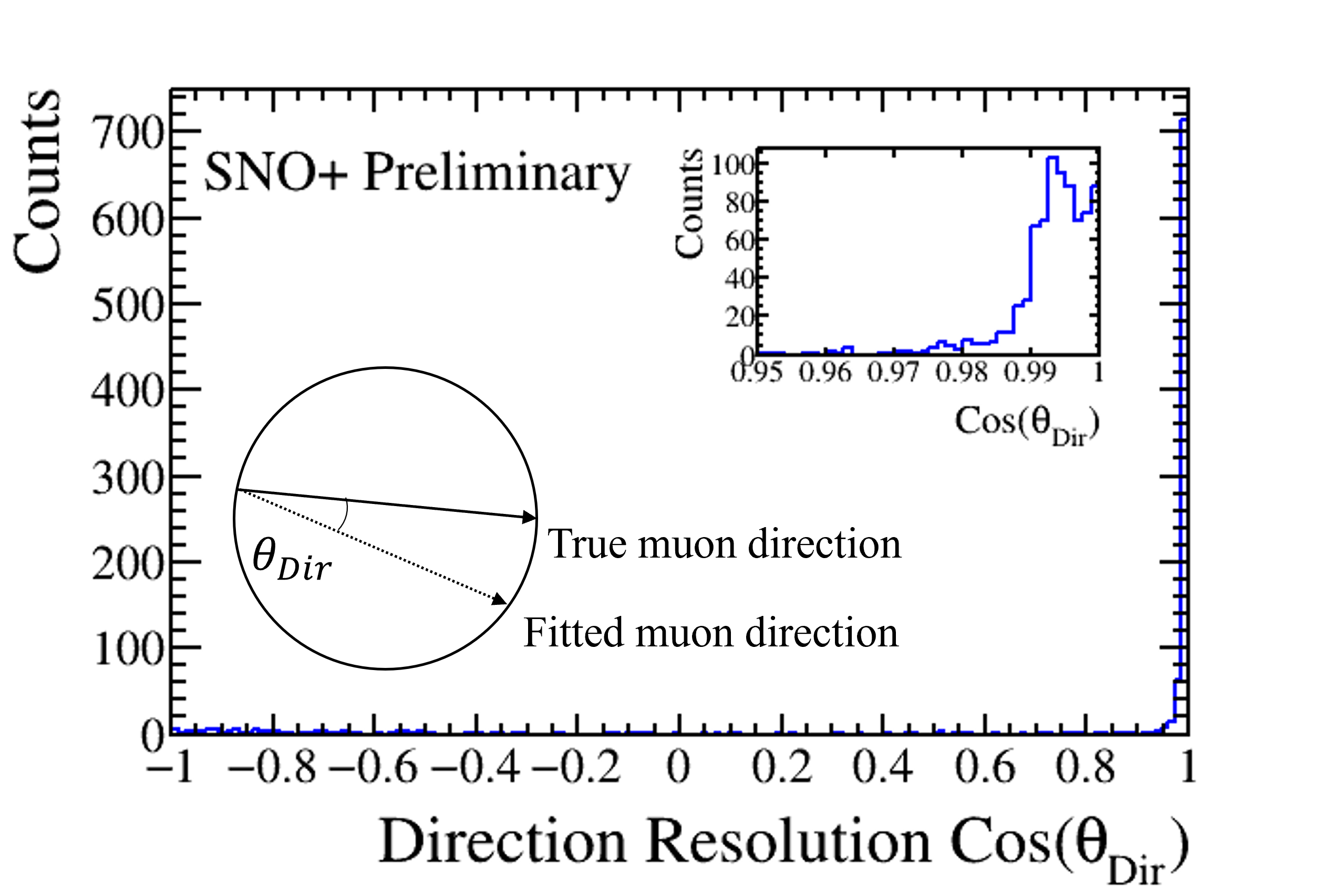}
\end{subfigure}
\caption{Distribution of $cos(\theta_{Pos})$ and $cos(\theta_{Dir})$ for simulated cosmic muons.}
\end{figure}

Figure 4 shows the results of $cos(\theta_{Pos})$ and $cos(\theta_{Dir})$ for the track reconstruction, where $\theta_{Pos}$ is the angle between the true and reconstructed muon entry position and $\theta_{Dir}$ is the angle between the the true and reconstructed muon direction. The position and direction resolution was calculated from the peak of these distributions and found to be $15^{\circ}$ and $7^{\circ}$ respectively.

\section{Conclusion}

This work demonstrates an effective muon track fitter for the scintillator phase of SNO+. This can be used to increase the livetime for all physics analyses in SNO+ by vetoing only events in a cylindrical volume around the muon track rather than a 20 s muon veto of the whole detector.

Initial tests with simulated muons in SNO+ show good agreement between the true and reconstructed tracks. Following the verification with the SNO+ scintillator data, the reconstrction algorithm will be merged into the SNO+ code.

\begin{Acknowledgements}
This work is supported by ASRIP, CIFAR, CFI, DF, DOE, ERC, FCT, FedNor, NSERC,
NSF, Ontario MRI, Queen’s University, STFC, and UC Berkeley, and have benefited
from services provided by EGI, GridPP, and Compute Canada. Jasmine Simms is supported by STFC (Science and Technologies Facilities Council). We thank Vale and
SNOLAB for their valuable support. 
\end{Acknowledgements}


\vspace{-4mm}
\begin{thebibliography}{9}
\vspace{-2mm}
\bibitem{...}
SNO+ Collaboration, Albenese V. et al, “The SNO+ Experiment”, J. Instrum.
16 (2021) P08059
\bibitem{...}
L. Nolan, “Cosmogenic Muon Induced Background Tagging in the SNO+ Detector”, PhD
thesis, Queen Mary University of London (2023)
\bibitem{...}
Q. Zhang, "Reconstruction of Straight Muon Tracks in the Detector" (2023)
\end{thebibliography}
\end{document}